\documentclass[aps,prl,reprint,superscriptaddress,showpacs]{revtex4-1}
\usepackage{graphicx}
\usepackage{graphics}
\usepackage{amsmath}
\usepackage{amssymb}
\usepackage{lmodern}
\usepackage{color}
\usepackage{nicefrac}
\usepackage[squaren]{SIunits}

\begin{document}

%\preprint{}

\title{Multi-photon nonclassical correlations in entangled squeezed vacuum states} %Title of paper



\author{Bhaskar Kanseri}
\email[]{bhaskar.kanseri@gmail.com}
%\thanks{}
\affiliation{Max-Planck Institute for the Science of Light, Guenther-Scharowsky Str. 1/Bldg. 24, 91058 Erlangen, Germany}
\author{Timur Iskhakov}
\affiliation{Max-Planck Institute for the Science of Light, Guenther-Scharowsky Str. 1/Bldg. 24, 91058 Erlangen, Germany}
\author{Georgy Rytikov}
\affiliation{Ivan Fedorov State University of Printing Arts, Moscow, Russia}
\author{Maria Chekhova}
\affiliation{Max-Planck Institute for the Science of Light, Guenther-Scharowsky Str. 1/Bldg. 24, 91058 Erlangen, Germany}
\affiliation{M. V. Lomonosov Moscow State University, 119992 GSP-2, Moscow, Russia}
\affiliation{University of Erlangen-Nurenberg, Staudtstrasse 7/B2, 91058 Erlangen, Germany}
\author{Gerd Leuchs}
\affiliation{Max-Planck Institute for the Science of Light, Guenther-Scharowsky Str. 1/Bldg. 24, 91058 Erlangen, Germany}
\affiliation{University of Erlangen-Nurenberg, Staudtstrasse 7/B2, 91058 Erlangen, Germany}

\date{\today}

\begin{abstract}
Photon-number correlation measurements are performed on bright squeezed vacuum states using a standard Bell-test setup, and quantum correlations are observed for conjugate polarization-frequency modes. We further test the entanglement witnesses for these states and demonstrate the violation of the separability criteria, which infers that all the macroscopic Bell states, containing typically $10^6$ photons per pulse, are polarization entangled. The study also reveals the symmetry of macroscopic Bell states with respect to local polarization transformations.
\end{abstract}

\maketitle %\maketitle must follow title, authors, abstract and \pacs

Contrary to single- and few-photon states, bright non-classical states of light contain large number of photons and thus resemble classical systems~\cite{De Martini,Iskhakov09}. It is quite interesting and demanding to investigate to what extent such states exhibit quantumness~\cite{Zeilinger}. Among available bright quantum states, here we deal with a four-mode (two polarization and two frequency) \textit{macroscopic squeezed vacuum} called the \textit{macroscopic Bell states} (MBS) \cite{Iskhakov11}. In recent years, these  states are in limelight owing to their stronger interaction properties with matter and with each other. The beauty of MBS is the manifestation of polarization squeezing either in one (the triplet states) or in all the three Stokes observables (the singlet state) despite the fact that all these states are unpolarized in the first order of the intensity \cite{Iskhakov11,Kanseri}.
The benefit of these quantum polarization states \cite{Bowen} over continuous-variable states is that they do not require a local oscillator for characterization and can be studied through quantum polarization tomography \cite{Kanseri,Bushev}, which is significantly simpler from technical point of view.

\begin{figure}[h]
\includegraphics{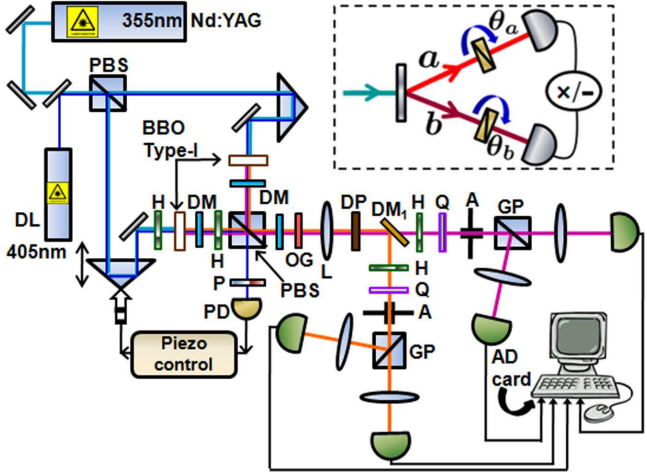}
\caption{\label{Fig.1}(color online) Experimental setup to test multiphoton quantum correlations and polarization entanglement. Picosecond UV pulses (Nd:YAG) pump two collinear, frequency non-degenerate optical parametric amplifiers (OPAs). The pump is cut off using dichroic mirrors (DM) and a long pass filter (OG) before recombining the OPA outputs on a polarizing beam splitter (PBS). Using an additional diode laser (DL) beam, which partially passes through all the components within the interferometer, and a feedback loop involving a piezo-actuator, a polarizer (P), and a photo-diode (PD), the interferometer is locked at $\pi$ phase. Making use of a dichroic plate (DP), the $\left|\Phi_{mac}^- \right\rangle$ and $\left|\Psi_{mac}^- \right\rangle$ states are prepared (see text for the details). Another dichroic mirror (DM$_1$) is used to reflect 635nm and transmit 805nm wavelengths and standard Stokes measurements are performed in each arm, with polarization transformations made by a half wave plate (H) and a quarter wave plate (Q). Inset: a typical Bell-test setup.}
\end{figure}

\textit{Entanglement} is a crucial trait of quantum mechanics \cite{Schroedinger}. Of late, it is treated as a resource, rather then a mystery and is one among the several signs of `quantumness' \cite{Horodecki}. In general, entanglement refers to the degree of correlation between the observables in conjugate subsystems, which surpasses any correlation allowed by the laws of classical physics \cite{Vedral}. Entanglement can manifest in many degrees of freedom, for example: in space, frequency, polarization etc.
%The polarization entanglement, in which the photon numbers for the conjugate polarization modes are correlated non-classically, may find potential applications in in \textcolor{blue}{quantum information \cite{Gisin,Braunstein}. In the past}, twin-beam multiphoton entanglement has been studied for the singlet state both in theory and in experiments \cite{Schroedinger, Horodecki,Braunstein,Reid,17,18,19,20}. However, to our knowledge, no experimental study in literature addresses the issue of bipartite multiphoton quantum correlations, which are very similar to the Bell-type correlations \cite{21}. At the same time, since all the MBS manifest non-classical features \cite{4,5}, it is also worth testing the MBS (both the triplet and the singlet states) for the polarization entanglement.
The most well-known example of polarization-entangled states are two-photon Bell states (see, for instance~\cite{Pan}). Their name originates from the fact that they violate the Bell inequality, setting the most profound boundary between the quantum and classical behavior. Bell's inequalities for two-photon states, derived from the assumptions that each part of a quantum system has (1) \emph{a priori} values of physical observables (\emph{hidden variables}) and (2) no nonlocal influence on the other part, have been violated in numerous experiments (see Ref.~\cite{Aspect} and, for a review,~\cite{Pan}).
%An important investigation for the non-classical squeezed states is to confirm whether such states are non-classically correlated. Among such quantum correlated states, a subset of states violates Bell inequalities and is generally referred to have ``Bell-type correlations". To demonstrate non-classical correlations, 
A typical scheme of such an experiment with photon pairs produced via spontaneous parametric down-conversion (SPDC)~\cite{Bell test} is shown in the inset to Fig.~\ref{Fig.1}. Photon pairs are emitted into separate beams $a,b$ with polarizations being uncertain, but whenever the photon in beam $a$ is found in a certain polarization state (which is revealed by passing it through a polarizing prism at angle $\theta_a$), its match in beam $b$ is found in a state with the corresponding polarization. This polarization correlation is registered by measuring the coincidence counting rate, which is 100\% modulated under the rotation of the polarizing prisms.

For instance, if the crystal emits the singlet Bell state $|\Psi^-\rangle$, the coincidence counting rate $R_c$ depends on the polarizer orientations as $R_c\propto\sin^2(\theta_a-\theta_b)$. In the case of a $|\Phi^-\rangle$ Bell state, the modulation is different, $R_c\propto\cos^2(\theta_a+\theta_b)$, but still has a 100\% visibility. The dependencies for the other Bell states are similar.

For all Bell states obtained via parametric down-conversion (PDC), the visibility remains 100\% only in the case of low gain, where the probability of multi-photon emission is low. %In agreement with this, the Bell inequalities are derived under the assumption that only a single photon pair is emitted and not more.
At high gain, there is a significant decrease in the visibility of coincidence count rates and thus the observed correlations do not appear non-classical. In particular, the Bell inequalities are no more violated.~\cite{DeMartini10}. 
 This is because coincidence counting rates are a measure of Glauber's correlation functions, and those acquire a huge background component at high-gain PDC. This, however, does not mean that quantum correlations no longer retain at high gain; it is just the wrong measurement that is used. If, instead of the correlation functions, one measures the variance of the difference photon number (for convenience normalized to the mean sum photon number, which yields the noise reduction factor, NRF), the correlations are revealed even at high gain and even in the presence of multiple modes. Namely, for properly selected polarization modes in beams $a,b$ the NRF is equal to zero for unity quantum efficiency. In recent years, this technique has been implemented successfully for investigating various features of MBS~\cite{Iskhakov11,Kanseri}. By measuring NRF in a standard Bell-test scheme we report the first experimental observation of multi-photon quantum correlations for all MBS, with the polarization modulation pronounced very distinctly and the NRF reaching values much below the shot-noise level. Our experiment does not include a Bell test; moreover, the corresponding Bell inequality has not yet been formulated, but we believe that our experimental results will stimulate the efforts on its derivation. Since all the triplet states can be transformed into one another using certain global polarization transformations, for brevity, we analyze only one of the triplet states, $\left|\Phi_{mac}^- \right\rangle$, and the singlet state $\left|\Psi_{mac}^- \right\rangle$. In earlier studies of entanglement, the emphasis was given to the singlet state only~\cite{Lamas-Linares,Eisenberg,Iskhakov12}. Now, we experimentally test the separability witnesses for the triplet states recently derived in Ref.\cite{Stobinska}, and demonstrate that not only the singlet MBS ~\cite{Iskhakov12} but also the triplet MBS are polarization entangled.

The polarization correlations of MBS are explicitly seen in their state vectors~\cite{Iskhakov12},
\begin{subequations}
\begin{equation}
\left|\Phi_{mac}^{\pm}\right\rangle=\sum_{n=0}^\infty \sum_{m=0}^n A_{nm}\left|n-m\right\rangle_{aH}\left|m\right\rangle_{aV}\left|n-m\right\rangle_{bH}\left|m\right\rangle_{bV},
\end{equation}
\begin{equation}
\left|\Psi_{mac}^\pm\right\rangle=\sum_{n=0}^\infty \sum_{m=0}^n A_{nm} \left|n-m\right\rangle_{aH}\left|m\right\rangle_{aV}\left|m\right\rangle_{bH}\left|n-m\right\rangle_{bV},
\end{equation}
\end{subequations}
with
\begin{equation}
A_{nm}=\frac{\text{sinh}^n\Gamma}{\text{cosh}^{n+2}\Gamma} (\pm1)^m,
\end{equation}
where
%the signs $\pm$ correspond to $\left|\Phi_{mac}^{\pm}\right\rangle$ and $\left|\Psi_{mac}^{\pm}\right\rangle$ states
$\Gamma$ is the parametric gain. Subscripts $a,b$ denote the frequency modes whereas $H,V$ denote the horizontal and the vertical polarization modes, respectively, and $n,m,n-m$ are photon numbers. From Eq. (1), we see that for the states $\left|\Phi_{mac}^{\pm}\right\rangle$, photon numbers in the same polarization but different frequency modes are exactly the same, whereas for $\left|\Psi_{mac}^{\pm}\right\rangle$ states, photon numbers in different polarization-frequency modes are identical. This, in an ideal situation, would lead to a complete reduction of noise in the difference of photon numbers for the corresponding pairs of modes. In a real experiment, the noise is reduced not to zero but to a value depending on the losses in the system. By applying to all four MBS the Heisenberg approach with an account for optical losses~\cite{Iskhakov11}, we have calculated the %noise in the photon-number difference for different pairs of modes. Table 1 lists the
NRF values for photon numbers ($N$) in modes $ai,bj$, $i,j=H,V$, defined as $\hbox{NRF}(N_{ai},N_{bj})\equiv\hbox{Var}(N_{ai}-N_{bj})/\langle N_{ai}+N_{bj}\rangle$. These values are listed in Table 1. The losses in the setup are taken into account by introducing the effective quantum efficiency $\eta$, and $n$ is the mean photon number per mode~\cite{DNK}. From Table 1, we see that at low gain ($n<<1$), the noise never exceeds much the shot-noise level,  NRF$\lesssim 1$. We also observe that at $\eta=1$, photon number fluctuations are completely suppressed (NRF=0) for orthogonally polarized modes in the case of $\left|\Psi_{mac}^{\pm}\right\rangle$ states and for similarly polarized modes in the case of $\left|\Phi_{mac}^{\pm}\right\rangle$ states.
\begin{table}[ht]
\caption{NRF for photon number difference corresponding to various combinations of polarization-frequency modes for macroscopic Bell states. } % title of Table
\centering  % used for centering table
\renewcommand{\tabcolsep}{0.18cm}
\begin{tabular}{c  c  c  c  c} % centered columns (4 columns)
\hline\hline
 %inserts double horizontal lines
Case & $\left|\Phi_{mac}^{+}\right\rangle$ & $\left|\Phi_{mac}^{-}\right\rangle$ & $\left|\Psi_{mac}^{+}\right\rangle$ & $\left|\Psi_{mac}^{-}\right\rangle$ \\ [0.5ex] % inserts table
%heading
\hline                  % inserts single horizontal line
NRF$(N_{a_H},N_{b_H})$ &$1-\eta$ & $1-\eta$ & $1+n\eta$ & $1+n\eta$\\ % inserting body of the table
NRF$(N_{a_H},N_{b_V})$ & $1+n\eta$& $1+n\eta$ & $1-\eta$ & $1-\eta$\\
NRF$(N_{a_V},N_{b_H})$ & $1+n\eta$ & $1+n\eta$  & $1-\eta$ & $1-\eta$\\
NRF$(N_{a_V},N_{b_V})$ & $1-\eta$ & $1-\eta$ & $1+n\eta$ & $1+n\eta$\\ [1ex]      % [1ex] adds vertical space
\hline %inserts single line
\end{tabular}
\label{table:1} % is used to refer this table in the text
\end{table}

The experimental setup is shown in Fig. 1. Using a Mach-Zehnder interferometer, the orthogonally polarized macroscopic squeezed vacuums generated in two 3mm BBO crystals are superimposed in one of the output PBS ports with a provision to control the relative phase between them (For more details of the MBS preparation, please see \cite{Iskhakov11,Kanseri}). Here, the optic axes of the crystals are both in the horizontal plane and tilted symmetrically with respect to the pump, so that the walk off is in the same direction and the down-converted beams fully overlap after the interferometer. To make the pump extraordinary for both crystals, half wave plates (H) are placed on each sides of one crystal (see Fig. 1) to rotate the polarization of the pump and the PDC by $90^0$. The relative phase between the two squeezed vacuums is made $\pi$, producing the $\left|\Phi_{mac}^- \right\rangle$ state. This state is converted into the $\left|\Psi_{mac}^- \right\rangle$ state by inserting a dichroic plate which introduces a $\pi$ difference in phase delays between the ordinary and extraordinary beams at the wavelengths $\lambda_a=635$nm and $\lambda_b=805$nm ~\cite{Iskhakov11}. A dichroic mirror (DM$_1$) is used to separate both these wavelengths, and after polarization transformations performed with phase plates, measurements are made in each arm using a p-i-n diode-based detector placed in each output port of a Glan prism (GP). The acquired signals scaling linearly with the number of the detected photons are time integrated and analyzed by an analog-digital (AD) card \cite{Iskhakov09}.

In the far field, for the best matching of transverse modes, two apertures (A) with the diameters $D_a=7.0$ mm, $D_b=8.91$ mm satisfying the condition $D_a/D_b=\lambda_a/\lambda_b$ \cite{Agafonov} were inserted in the two beams, at the back focal plane of a 20 cm lens (L) having the crystals at its front focal plane (Fig. 1). In each pulse, the average number of measured photons was approximately $10^6$. The shot noise level (SNL) was obtained by calibrating the combination of four detectors using an attenuated second harmonic of the laser (532 nm) and the balanced detection scheme \cite{Iskhakov09}.
\begin{figure}
\includegraphics{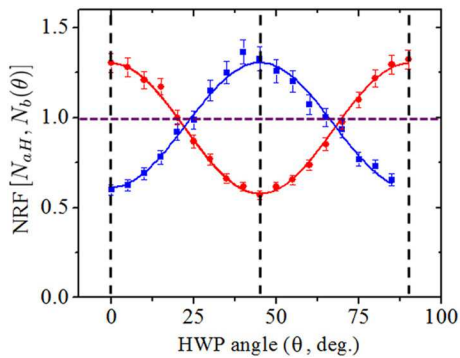}
\caption{\label{Fig.2}(color online) NRF for photon numbers registered in the transmittance port of the Glan prism in the $635$nm arm and the reflectance port of the other Glan prism vs the HWP orientation in the $805$nm arm. Red circles represent the triplet $\left|\Phi_{mac}^- \right\rangle$ state whereas blue squares show the singlet $\left|\Psi_{mac}^- \right\rangle$ state measurement. The lines show the theoretical fits using Eq. (3). NRF$=1$ corresponds to the shot noise level.}
\end{figure}

The first measurement results, as shown in Fig. 2, demonstrate photon-number correlations between conjugate polarization-frequency modes for the triplet and singlet states. The NRF for the photon numbers detected at different output ports of the two Glan prisms is observed as a function of the orientation of the half-wave plate (HWP) placed in the 805nm arm. For $\left|\Phi_{mac}^- \right\rangle$ state, the fluctuations are minimal when the HWP is at $45^0$, corresponding to the measurement for the same polarization modes. Whereas for the singlet state, the fluctuations are suppressed for $0^0$ and $90^0$ orientation of HWP, which represents the measurement of orthogonal polarization modes. For both the states, the maximum squeezing was obtained nearly 2.6dB below the shot noise level. The experimental curves depicted in Fig. \ref{Fig.2} are fitted using the theoretical dependence obtained by calculating in the Heisenberg picture the action of the HWP on the photon-number operators and their second moments, which gives
\begin{equation}
\text{NRF}[N_{aH},N_b(\theta)]=1+\eta[(1+n) \text{cos}^22\theta-1],
\end{equation}
for the $\left|\Psi_{mac}^- \right\rangle$ state, and a similar dependence with the cosine replaced by sine for the $\left|\Phi_{mac}^- \right\rangle$ state. The fitting parameters are the overall quantum efficiency $\eta$ and the mean photon number per mode $n=\text{sinh}^2\Gamma$. They were found to be $\eta =0.40\pm0.06$ and $n=0.8\pm0.2$. Following the symmetry properties of the Hamiltonians for the MBS, it would be easy and straightforward to show the similar quantum correlations for other triplet states. These observations confirm the existence of multiphoton quantum correlations in all MBS. Note that the visibility of the observed modulation is about 40\%, which is much higher than the value coincidence counting would give in this case. At the same time, noise reduction by 43\% is observed, which is a vivid demonstration of nonclassical behavior.

We would like to stress once again that the observed correlations do not enable a Bell test as Bell inequalities based on the variances of photon-number differences have not been constructed yet. Their derivation will be an important future step in the study of macroscopic entanglement.

\begin{figure*}
\includegraphics{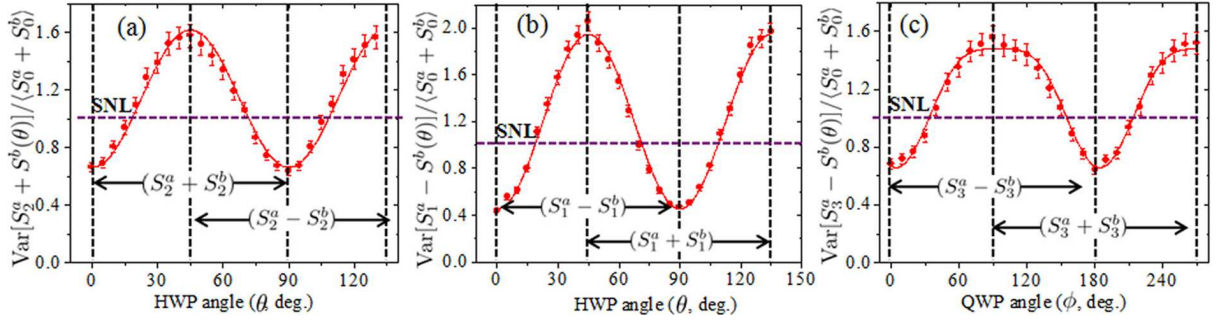}
\caption{\label{Fig.3} Effect of phase plate rotations on the normalized variances of the combinations of the Stokes observables for the triplet $\left|\Phi_{mac}^- \right\rangle$ state. (a) HWP(635) was at $22.5^0$ and HWP(805) was rotated as $22.5^0+\theta$, (b) HWP(635) was at $45^0$ and HWP(805) was rotated as $45^0+\theta$, and (c) QWP(635) was at $45^0$ and QWP(805) was rotated as $45^0+\phi$. The theoretical fits made using Eqs. (7) and (8) are shown by red lines.}
\end{figure*}

Entanglement witnesses are the observables used to `detect' entanglement (qualitatively) but not to `measure' it (quantitatively). The entanglement witness can be expressed for the triplet state $\left|\Phi_{mac}^- \right\rangle$ as \cite{Stobinska}
\begin{equation}
\begin{split}
W_{\left|\Phi_{mac}^- \right\rangle}=\text{Var}(S_1^a-S_1^b)+\text{Var}(S_2^a+S_2^b)\\
+\text{Var}(S_3^a-S_3^b)-2\langle S_0^a+S_0^b \rangle,
\end{split}
\end{equation}
and for the singlet state $\left|\Psi_{mac}^- \right\rangle$ as
\begin{equation}
\begin{split}
W_{\left|\Psi_{mac}^- \right\rangle}=\text{Var}(S_1^a+S_1^b)+\text{Var}(S_2^a+S_2^b)\\
+\text{Var}(S_3^a+S_3^b)-2\langle S_0^a+S_0^b \rangle.
\end{split}
\end{equation}
The negative value of these witnesses is a sufficient condition for the triplet and the singlet states to be non-separable. This is equivalent to the conditions
\begin{subequations}
\begin{equation}
[\text{Var}(S_1^a-S_1^b)+\text{Var}(S_2^a+S_2^b)+\text{Var}(S_3^a-S_3^b)]/\langle S_0^a+S_0^b\rangle< 2,
\end{equation}
\begin{equation}
[\text{Var}(S_1^a+S_1^b)+\text{Var}(S_2^a+S_2^b)+\text{Var}(S_3^a+S_3^b)]/\langle S_0^a+S_0^b\rangle< 2.
\end{equation}
\end{subequations}

%\textcolor{blue}{NRF for the sum and the difference Stokes observables in modes $a$ and $b$ can be defined as %$\text{NRF}(S_n^a,S_n^b)_\pm=\text{Var}(S_n^a\pm S_n^b)/\langle S_0^a+S_0^b\rangle$ where $n=1,2,3$ and $S_0$ is the %zeroth Stokes observable.

The non-separability witnesses have been tested for the MBS by measuring the individual terms of Eqs. (6). For the triplet state $\left|\Phi_{mac}^- \right\rangle$, Fig. 3 shows the dependence of the normalized variances for various observables~\cite{Stokes} on the local polarization transformation (performed only on one frequency mode). The measured squeezing was 2dB, 3.6dB and 2dB for the observables $S_2^a+S_2^b$, $S_1^a-S_1^b$, and $S_3^a-S_3^b$, respectively. From these measurements, the left-hand side (LHS) of inequality (6a) gives a value of $1.75\pm 0.03$, which is well below its right-hand side (RHS). Hence, the triplet state under test is polarization entangled. The plots demonstrate the behavior of the macroscopic triplet state $\left|\Phi_{mac}^- \right\rangle$ under local polarization transformations similar to the one obtained for two-photon triplet state $\left|\Phi^- \right\rangle$ in coincidence measurements \cite{AVB}.

\begin{figure}
\includegraphics{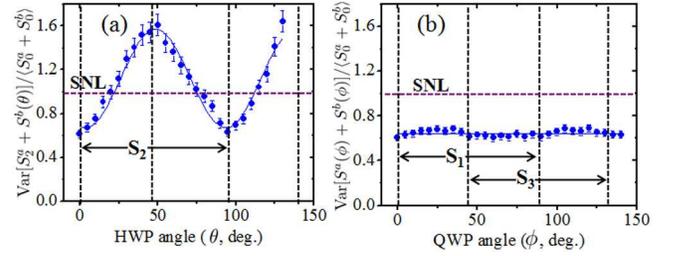}
\caption{\label{Fig.4} Effect of polarization transformations (PT) on the normalized variances of the combinations of Stokes observables for the singlet $\left|\Psi_{mac}^- \right\rangle$ state. (a) Local PT when HWP(635) at $22.5^0$ and HWP(805) was rotated as $22.5^0+\theta$. Blue line shows theoretical fit using Eq. (\ref{Psi}). (b) Global PT when QWPs in both frequency modes were rotated in the same (clockwise) direction.}
\end{figure}

Similar measurements have been carried out for the singlet state $\left|\Psi_{mac}^- \right\rangle$. Figure 4(a) shows the dependence of the normalized variance of the measured observable \cite{Stokes} on the orientation of HWP in one arm. This dependence, again, shows the behavior of the macroscopic singlet state under a local polarization transformation. On the other hand, the state is invariant to global polarization transformations (the same polarization transformations for both frequency modes), as shown in part (b). The squeezing obtained for the combinations $S_1^a+S_1^b$, $S_2^a+S_2^b$, $S_3^a+S_3^b$ was 2.2dB, 2.1dB and 2.2dB, respectively. The LHS of Eq. (6b) results in a value of $1.82\pm 0.03$, which is less than the RHS, i.e. 2, demonstrating that the singlet state is non-separable (entangled).

The plots given in Figs. 3 and 4 are fitted with the dependencies calculated by considering the effect of plate orientations on the second-order moments of photon-number operators. For the triplet state, the dependence on the HWP orientations $\theta_{a,b}$ in channels $a,b$ is expected to be
\begin{equation}
\frac{\text{Var}[S^a(\theta_a)\pm S^b(\theta_b)]}{\langle S_0^a+ S_0^b\rangle}=1+\eta[n\pm(1+n)\text{cos}4(\theta_a+\theta_b)],
\label{Phi,theta}
\end{equation}
which fits well the experimental points of Fig. 3(a), (b). At the same time, if the Stokes observable $S_3$ is measured in channel $a$ and a quarter wave plate (QWP) is rotated by angle $\phi$ in channel $b$, the variance of the difference Stokes observable is given by
\begin{equation}
\frac{\text{Var}[S_3^a-S^b(\phi)]}{\langle S_0^a+ S_0^b\rangle}=1+n\eta-\frac{(1+n)\eta}{4}[1-4\text{cos}2\phi- \text{cos}4\phi].
\label{Phi,phi}
\end{equation}
The resulting fit is in a good agreement with the experimental results of Fig. 3(c). Finally, the dependence in Fig.~4(a) is described by the relation
\begin{equation}
\frac{\text{Var}[S^a(\theta_a)\pm S^b(\theta_b)]}{\langle S_0^a+ S_0^b\rangle}=1+\eta[n\mp(1+n)\text{cos}4(\theta_a-\theta_b)].
\label{Psi}
\end{equation}
Again, the theoretical dependence fits well the experimental points. The asymmetry in the curves shown in Figs. 3 and 4(a) with respect to the shot noise level can be explained from Eqs. (\ref{Phi,theta},\ref{Phi,phi},\ref{Psi}): the squeezing depends only on $\eta$ whereas the anti-squeezing depends on both $n$ and $\eta$. This is why with increasing gain, the anti-squeezing becomes more pronounced than the squeezing.

In conclusion, multi-photon quantum correlations in conjugate polarization modes have been observed for the macroscopic Bell states, prepared via combining two orthogonally polarized frequency nondegenerate bright squeezed vacuums generated from coherent pump beams. By measuring normalized variance of the difference photon numbers, correlations resembling those obtained in Bell tests were observed with visibility and noise reduction as high as 40\% and 43\%, respectively. Macroscopic polarization entanglement was probed for these bright states via testing the entanglement witnesses. For the singlet and triplet investigated MBS, the separability condition was violated by five and nine standard deviations, respectively. In addition, we have demonstrated the behavior of the MBS with respect to local polarization transformations. All these results were obtained for light containing about $0.8$ photons per mode but more than $10^6$ photons per pulse. With expected applications in quantum key distribution, quantum memories, and quantum gates,
%\begin{verbatim*}
%\end{verbatim*}
 these bright entangled states may serve as a building block for future quantum networks.

We are thankful to Magda Stobinska and Pavel Sekatsky for fruitful discussions. T.Sh.I. acknowledges funding from the Alexander von Humboldt Foundation. The research leading to these results has received
funding from the European Union Seventh Framework Programme under grant agreement No 308803 (project
BRISQ2) and partial support from the Russian Foundation for Basic Research, grant \#12-02-00965a.



\begin{thebibliography}{99}
\bibitem{Stokes} The observable here is $S_i^a\pm S^b(\theta)$ for ($i=1,2,3$), where $\theta$ (or $\phi$) is the orientation of a HWP (or QWP) placed in b mode. This angle determines the polarization transformation and thus the measured Stokes observable.
\end{thebibliography}
\end{document}